\documentclass{emulateapj}
\usepackage{apjfonts}

\newcommand{\deltavec}{\mbox{\boldmath $\delta$}}

\lefthead{HAN \& KIM}
\righthead{PLANETARY-LENSING SIGNALS OF A FINITE SOURCE}

\begin{document}
\title{Planetary Lensing Signals of High-Magnification Events\\ 
under Severe Finite-Source Effect}

\author{Cheongho Han and Doeon Kim}
\affil{
Program of Brain Korea 21,
Department of Physics,\\
Chungbuk National University, Chongju 361-763, Korea;
cheongho@astroph.chungbuk.ac.kr}

\submitted{Submitted to The Astrophysical Journal}

\begin{abstract}
We investigate the effect of a finite source on 
the  planetary-lensing signals of high-magnification 
events.  From this, we find that the dependency of 
the finite-source effect on the caustic shape is 
weak and perturbations survive even when the source 
is substantially bigger than the caustic.  
Specifically, we find that perturbations with 
fractional magnification excess $\geq 5\%$ 
survive when the source star is roughly 4 times 
bigger than the caustic.  We also find characteristic 
features that commonly appear in the perturbation 
patterns of planetary lens systems affected by 
severe finite-source effect and thus can be used 
for the diagnosis of the existence of a companion.  
These features form in and around a circle with 
its center located at the caustic center and a 
radius corresponding to that of the source star.  
The light curve of an event where the source 
crosses these features will exhibit a distinctive 
signal that is characterized by short-duration 
perturbations of either positive or negative 
excess and a flat residual region between these 
short-duration perturbations.
\end{abstract}


% ==================================================================

\section{Introduction}

Since the first discovery in 2004, eight planets have been 
detected by using the microlensing method \citep{bond04, 
udalski05, beaulieu06, gould06, gaudi08, bennett08, dong08}.  
The microlensing method is important in various aspects of 
exoplanet studies.  First, due to its high sensitivity to 
planets located in the outer region of planetary systems 
beyond the snow line, the microlensing method is complementary 
to other methods such as the radial velocity and the transit 
methods that are sensitive to planets orbiting close to their 
host stars.  In addition, the sensitivity of the microlensing 
method extends to very low-mass planets and Earth-mass planets 
can be detected from ground-based observations.  Furthermore, 
it is the only proposed method that can detect free-floating 
planets \citep{bennett02, han06b} that are thought to be kicked 
out from planetary systems.  The detection rate of microlensing 
planets is rapidly increasing and at least five additional 
planet candidates were detected during the 2007 and 2008 
observation seasons (A.\ Gould 2008, private communication).

The microlensing signal of a planet is a perturbation to a 
smooth standard light curve of a primary-induced lensing 
event occurring on a background star \citep{mao91, gould92}.  
The duration of the perturbation is short: several hours for 
an Earth-mass planet and several days even for a gas-giant 
planet.  Thus it is difficult to detect planetary signals 
from microlensing survey observations where stars are monitored 
on a roughly nightly basis.  Currently, the observational 
frequency required for planet detections is achieved by 
employing an early-warning system and follow-up observations, 
where the early-warning system (OGLE: Udalski et al.\ 1994; 
MOA: Bond et al.\ 2001) enables to issue alerts of ongoing 
events by analyzing data from survey observations real time 
and follow-up observations (PLANET: Kubas et al.\ 2008; 
Micro-FUN: Dong et al.\ 2006) are focused on these alerted 
events.  However, the limited number of telescopes available 
for follow-up observations restricts the number of events 
that can be followed up at any given time and thus priority 
is given to events which will maximize the planet detection 
probability.  Currently, the highest priority is given to 
high-magnification events \citep{bond02, yoo04}.  These 
events have high intrinsic planet detection efficiency 
because the source trajectories always pass close to the 
perturbation region around the central caustic induced by 
the planet \citep{griest98}.

Despite the high chance of perturbation, however, it is 
often thought that detecting low-mass planets through the 
channel of high-magnification events would be difficult.  
This thought is based on the fact that the central caustic 
induced by a low-mass planet is usually smaller than the
source star and thus the planetary signal would be greatly 
washed out by severe finite-source effect \citep{bennett96}.  
However, it might be that perturbations persist despite 
the finite-source effect and could still be detected thanks 
to the high photometry precision achieved by the enhanced 
brightness of the highly magnified source star.  In this 
paper, we test this possibility by investigating how the 
pattern of central planetary perturbations is affected by 
the finite-source effect.

The paper is organized as follows.  In \S\ 2, we briefly 
describe the physical properties of central caustics.  
In \S\ 3, we investigate the effect of a source size 
on the perturbation pattern.   For this, we construct 
maps of perturbation pattern for planetary systems with 
various caustic/source size ratios and caustic shapes.  
Based on these maps, we search for characteristic 
features that may be used to identify the existence of 
planets.  We summarize the results and conclude in \S\ 4.

\section{Central Caustic}

A planetary lens corresponds to the case of a binary lens 
with a very low-mass companion.  One important characteristic 
of binary lensing that differentiates it from those of single 
lensing is the formation of caustics, which represent the 
positions on the source plane where the lensing magnification 
of a point source becomes infinite.  Then, the light curve 
of a lensing event resulting from the source trajectory 
approaching or crossing the caustic can be dramatically 
different from the smooth and symmetric light curve of a 
single-lensing event.  The set of caustics form closed 
curves, each of which is composed of concave curves (fold 
caustic) that meet at points (cusps).  For a planetary case, 
there exist two sets of caustics.  A single `central caustic' 
is located close to the primary lens and the other one or 
two `planetary caustics' are located away from the primary.

Compared to size of the planetary caustic, the size of the 
central caustic is much smaller.  When the size of the central 
caustic is measured as the separation between the two cusps 
located along the primary-planet axis, it is represented by 
\citep{chung05}
\begin{equation}
\Delta\xi_{cc}\sim {4q\over (s-s^{-1})^2} \propto q,
\label{eq1}
\end{equation}
where $q$ represents the planet/primary mass ratio and $s$ 
is the primary-planet separation normalized by the Einstein 
radius corresponding to the total mass of the planetary 
system.  On the other hand, the size of the planetary caustic 
is\footnote{We note that the expression in eq.~(\ref{eq2}) is
for caustics of planetary lenses with $s>1$.  The planetary 
caustic for the case of $s<1$ has a different dependency on 
$s$, but it has the same dependency on $q$, i.e.\ $\propto 
q^{1/2}$ \citep{han06a}.} \citep{han06a}
\begin{equation}
\Delta\xi_{pc}\sim {4q^{1/2}\over s(s^2-1)^{1/2}} \propto q^{1/2}.
\label{eq2}
\end{equation}
Then, the size ratio between the two types of caustic is
\begin{equation}
{\Delta\xi_{cc}\over \Delta\xi_{pc}}=
{q^{1/2}\over (1-s^{-2})^{3/2}} \propto q^{1/2}.
\label{eq3}
\end{equation}
Considering that the size ratio is proportional to $q^{1/2}$ 
and the mass ratio is very small for planetary lenses, the 
central caustic is much smaller than the planetary caustic.  
In addition, the size ratio becomes even smaller for lower 
mass planets.

The small size of a central caustic implies that the planetary 
signal induced by the central caustic is more likely to be 
affected by larger finite-source effect.  The finite-source 
effect smears out the planetary signal because the magnification 
of a finite source corresponds to the magnification averaged over 
the source-star surface, i.e.\ 
\begin{equation}
A={
\int_0^{\rho_\star} I(r)A_p(|{\bf r}-{\bf r}_0|) r dr
\over
\int_0^{\rho_\star} I(r) r dr
},
\label{eq4}
\end{equation}
where ${\bf r}_0$ is the displacement vector of the source center
with respect to the lens, ${\bf r}$ is the vector to a position
on the source star surface from the center of the source star, 
$I(r)$ represents the surface brightness profile of the source
star, $A_p$ is the point-source magnification, and $\rho_\star$ 
represents the source radius normalized by the Einstein radius.

While the size of the central caustic depends both on the mass 
ratio and the separation, the shape of the caustic is solely 
dependent on the separation.  When the shape is quantified as 
the vertical/horizontal width ratio and the vertical width 
$\Delta\eta_{cc}$ is measured as the separation between the two 
off-axis cusps, the width ratio is related to the planetary 
separation by \citep{chung05}
\begin{equation}
{\Delta\eta_{cc}\over\Delta\xi_{cc}}=
{(s-s^{-1})^2|\sin^3\phi|
\over
(s+s^{-1}-2\cos\phi)^2},
\label{eq5}
\end{equation}
where 
\begin{equation}
\phi = \cos^{-1} 
\left\{
{3\over 4}(s+s^{-1})
\left[ 1-\sqrt{1-{32\over 9}
\left( s+s^{-1}\right)^{-2}}\right]
\right\}.
\label{eq6}
\end{equation}
For a given mass ratio, a pair of central caustics with separations
$s$ and $s^{-1}$ are identical to the first order of approximation.

% Figure 1 ----------------------------------------------------
\begin{figure*}[htb]
\epsscale{0.9}
\plotone{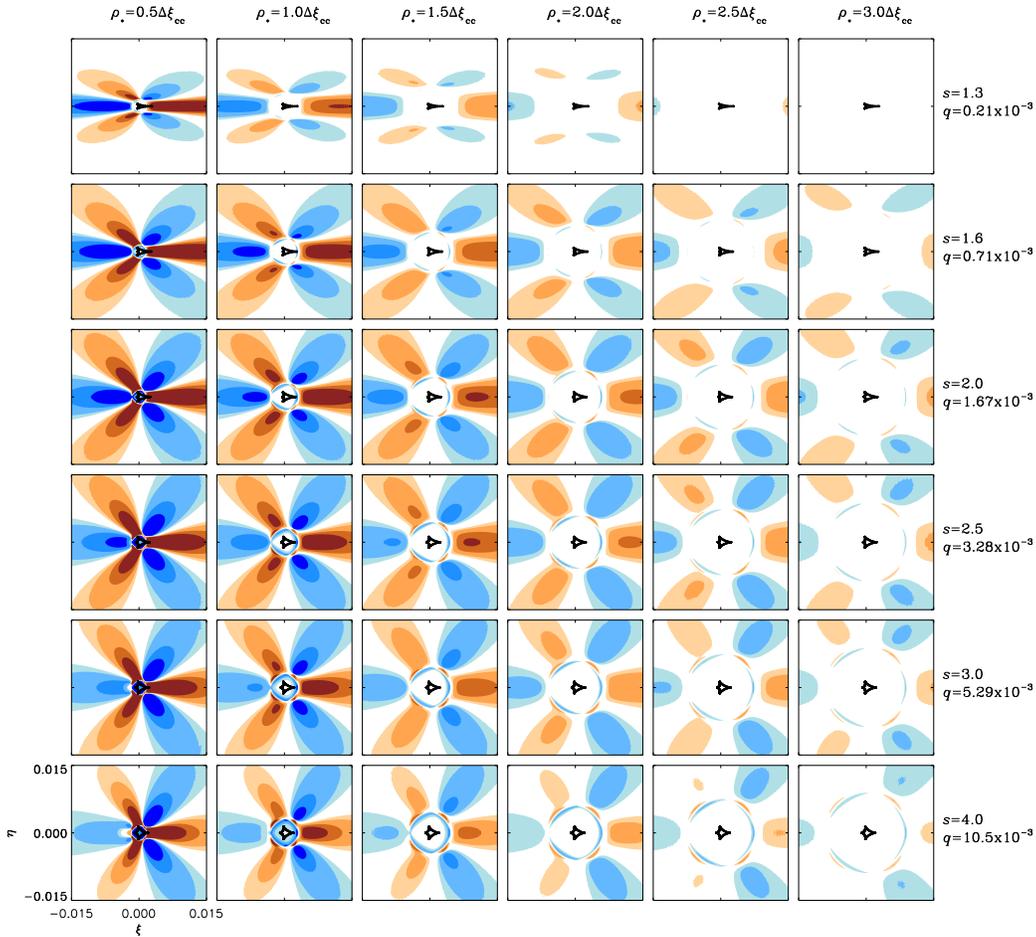}
\caption{\label{fig:one}
Maps of magnification excess around the region of central 
caustics with various shapes and caustic/source size ratios.  
In the figure, the panels in each column show the pattern 
variation depending on the caustic shape, while the panels in 
each row shows the variation depending on the caustic/source 
size ratio.  In each map, the regions with brown and blue-tone 
colors represent the areas where the planetary lensing 
magnification is higher $(\epsilon > 0)$ and lower $(\epsilon 
< 0)$ than the single-lensing magnification, respectively.  
For each tone, the color changes into dark scales when the 
excess is $|\epsilon| \geq 2\%$, 3\%, 5\%, and 7\%, 
respectively.  The coordinates $(\xi,\eta)$ are aligned so 
that $\xi$ axis matches the star-planet axis and the center 
is located at the effective position of the primary lens 
around which the single-lensing magnification best describes 
that of the planetary lensing.  The planet is located on the 
right.  All lengths are in units of the Einstein radius 
corresponding to the total mass of the lens system.  The 
relation between the source and caustic sizes is marked on 
the top of each column.  The primary-planet separation and 
the mass ratio are marked on the right side of each row.
}\end{figure*}
% -------------------------------------------------------------

\section{Perturbation Pattern}

The pattern of planetary perturbation is determined by two 
major factors.  The first factor is the caustic shape and it 
determines the basic pattern of the perturbation.  The second 
factor is the size ratio between the caustic and the source 
star and it characterizes how the pattern is deformed by the 
finite-source effect.

To see how a source size affects the pattern of central 
perturbations, we construct maps of {\it magnification excess}
in the region around central caustics with various shapes and 
sizes.  The magnification excess is defined by 
\begin{equation}
\epsilon={A-A_0 \over A_0},
\label{eq7}
\end{equation}
where $A$ and $A_0$ represent the lensing magnifications with 
without the planet, respectively.

Figure~\ref{fig:one} show the constructed maps.  In the figure, 
the panels in each column show the pattern variation depending 
on the caustic shape, while the panels in each row show the 
variation depending on the caustic/source size ratio.  We note 
that all caustics have an identical size as measured by the 
horizontal width $\Delta\xi_{cc}$.  In each map, the regions 
with brown and blue-tone colors represent the areas where the 
planetary-lensing magnification is higher $(\epsilon>0)$ and 
lower $(\epsilon<0)$ than the single-lensing magnification, 
respectively.  For each tone, the color changes into dark scales 
when the excess is $|\epsilon|\geq 2\%$, 3\%, 5\%, and 7\%, 
respectively.  The coordinates $(\xi,\eta)$ are aligned so that 
$\xi$ axis matches the primary-planet axis and the center is 
located at the effective position of the primary lens.  The 
effective lens position represents the location of a single 
lens around which the magnification best describes that of 
the planetary lensing and it approximately corresponds to the 
center of the caustic.  For a planetary case, the effective 
position has an offset from the original position of the 
primary of 
\begin{equation}
\deltavec = 
\cases{
 {\bf s}^{-1}q/(1+q)    & for $s>1$, \cr
-{\bf s}[(1+q)^{-1}-1] & for $s<1$,  \cr
}
\label{eq8}
\end{equation}
where the sign is positive when the offset vector is directed 
toward the planet.  In the map, the planet is located on the 
right and all lengths are normalized by the Einstein radius.  
We take the finite-source effect into consideration by 
modelling the source brightness profile as \begin{equation}
{I(\theta)\over I_0}=
1-\Gamma\left( 1-{3\over 2}\cos\theta\right)
-\Lambda \left( 1-{5\over 4}\cos^{1/2}\theta\right),
\label{eq9}
\end{equation}
where $\theta$ is the angle between the normal direction 
to the source-star surface and the line of sight.  We adopt 
a linear and a square-root limb-darkening coefficients of 
$(\Gamma, \Lambda)=(-0.46,1.11)$.  The relation between the 
source and caustic sizes is marked on the top of each column.  
The primary-planet separation and the planet/primary mass 
ratio of the planetary system are marked on the right side 
of each row.

Considering that the caustic shape is characterized solely by 
the primary-planet separation and for a fixed planet/primary 
mass ratio the caustic size is linearly proportional to the 
mass ratio,  one can picture excess maps for other cases 
of planetary systems based on the presented maps in 
Figure~\ref{fig:one}.  For example, a planetary system with 
$(s,q)=(2.0,1.67\times 10^{-4})$ has a similar perturbation 
pattern to that of the system with $(s,q)=(2.0,1.67\times 
10^{-3})$. The only major difference is that the scale of 
the map decreases by a factor 10 because the caustic becomes 
smaller. We also note that the perturbation patterns of a pair 
of planetary systems with $s$ and $s^{-1}$ are identical and 
thus one can infer perturbation patterns for the cases of 
planetary systems with $s<1$ from those presented in the figure.

From the analysis of the maps, we find that the dependency 
of the finite-source effect on the caustic shape is weak 
and perturbations persist even when the source is substantially 
bigger than the caustic.  Specifically, we find that 
perturbations with $\epsilon\geq 5\%$ survive when the source 
star (as measured by its diameter) is roughly 4 times bigger 
than the caustic, i.e.\ $\rho_\star\sim 2\Delta \xi_{cc}$.  
Based on this, \citet{han08} derived an analytic expression 
for the optimal range of planetary separations (lensing zone) 
detectable through the channel of high-magnification events of 
\begin{equation}
\left\vert
\sqrt{{2q\over \rho_\star}} - \sqrt{{2q\over \rho_\star}+1}
\right\vert
\lesssim s \lesssim
\sqrt{{2q\over \rho_\star}} + \sqrt{{2q\over \rho_\star}+1}.
\label{eq10}
\end{equation}
Then, the lensing zone of central perturbations (central 
lensing zone) is different for planets with different mass 
ratios unlike the fixed range of the classical lensing zone 
($0.6\lesssim s \lesssim 1.6$) regardless of the mass ratio 
\citep{wambsganss97, griest98}.  According to this expression, 
the central lensing zone is larger than the classical lensing 
zone for planets with $q\gtrsim 3\times 10^{-4}$ despite the 
smaller size of the central caustic than the size of the 
planetary caustic.

% Figure 2 ----------------------------------------------------
\begin{figure}[t]
\epsscale{1.2}
\plotone{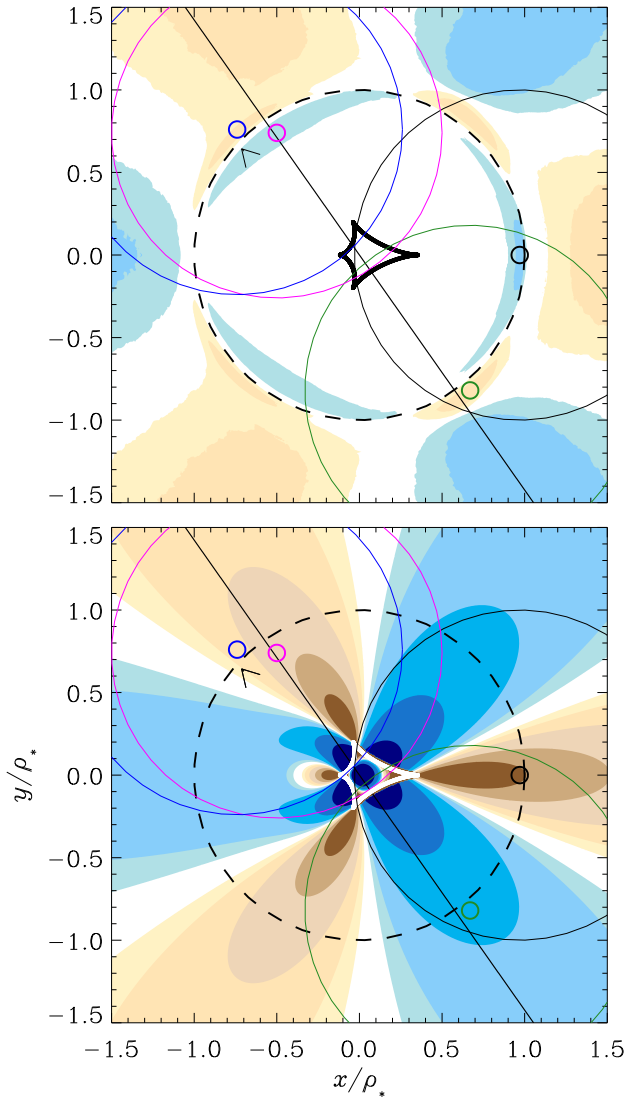}
\caption{\label{fig:two}
Excess maps of a planetary lens with (upper panel) and 
without (lower panel) finite-source effect.  Notations are 
similar to those of Fig.~\ref{fig:one} except two things.  
The first difference is that all lengths are scaled by the 
source radius not by the Einstein radius.  Another difference 
is that the color scale changes at the levels of  $|\epsilon|
= 2\%$, 4\%, 8\%, 16,\%, and 32\%, respectively.  Five circles 
are drawn in each map.  All circles have a common radius 
corresponding to the source radius.  The center of the black 
dashed circle is located at the effective primary position.  
The centers of other circles are located at the mid-points of 
the individual arc-shaped perturbation regions.  The planetary 
lens parameters are $s=0.38$ and $q=2.27\times 10^{-3}$
and the source radius is $\rho_\star=3.14\times 10^{-3}$.
}\end{figure}
% -------------------------------------------------------------

From the maps, we also find interesting features in the 
region near the boundary and inside of a circle with its 
center located at the caustic center and a radius corresponding 
to that of a source star.  These features are the localized 
arc-shaped perturbation regions with either positive or 
negative excess located at the edge of the circle and the 
region inside the circle with a very small magnification 
excess.  We find that these features commonly appear in 
the perturbation patterns affected by severe finite-source 
effect.

For close investigation of these features, we construct a 
separate set of excess maps of a planetary lens with and 
without the finite-source effect.  These maps are presented 
in Figure~\ref{fig:two}.  In the map, all lengths are scaled 
by the source radius to better show the location of the 
features in units of the source radius.  Another difference 
of the maps from those presented in Figure~\ref{fig:one} is 
that the color scale changes at the levels of  $|\epsilon|= 
2\%$, 4\%, 8\%, 16,\%, and 32\%, respectively, to emphasize 
the region of very high excesses.  Five circles are drawn on 
each map.  All circles have a common radius corresponding 
to the source radius.  The center of the black dashed circle 
is located at the caustic center.  The centers of other 
circles are located at the mid-points of the individual 
arc-shaped perturbation regions.

Close inspection of the pattern leads us to find the following 
tendencies.  First, we find that the very small excess in 
the region inside the dashed circle is caused by the cancellation 
of the positive and negative excesses.  When the center of the 
source star is located within this area, the amount of positive 
and negative excesses balances each other, resulting in a very 
small excess.  However, when the source center is located 
outside of this area, this balance breaks down and perturbation 
features show up.

% Figure 3 ----------------------------------------------------
\begin{figure}[t]
\epsscale{1.2}
\plotone{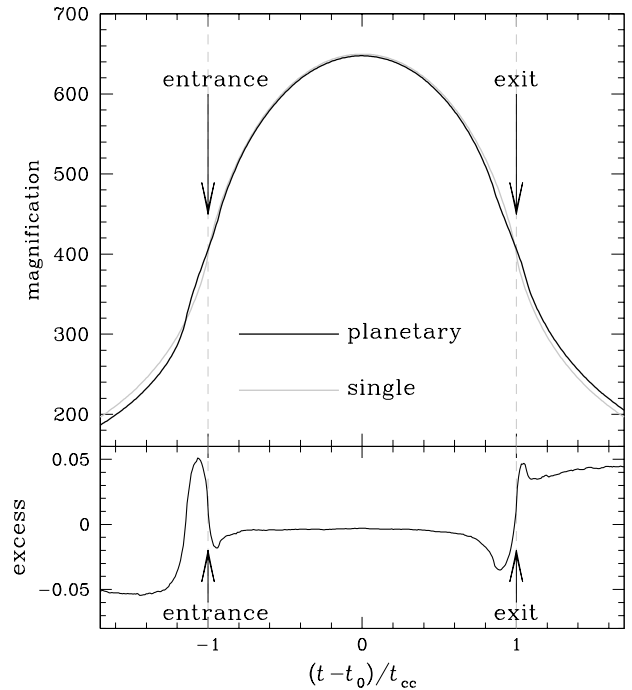}
\caption{\label{fig:three}
A characteristic planetary signal in the light curve of a 
high-magnification event affected by severe finite-source 
effect. The signal is characterized by short-duration 
perturbations of either positive or negative excess at the 
moments when the center of the central caustic enters into 
and exits from the source star and a flat residual between 
the short-duration perturbations. 
}\end{figure}
% -------------------------------------------------------------

Second, we find that the arc-shaped perturbation features are 
caused by the localized regions of very large excesses around 
the caustic.  Such regions include the regions just outside 
the cusps with very large positive excesses and the regions 
just outside the fold caustics with very large negative 
excesses.  Another region of a very large negative excess 
is the region inside the caustic.  The localized arc-shaped 
perturbation regions form at the source position where the 
source star encompasses some of these strong perturbation 
regions and the resulting overall excess is seriously unbalanced.  
For example, when the source star is located at the position 
of the blue circle, it encompasses two large positive-excess 
regions around two cusps but only a single large negative-excess 
region between the cusps.  As a result, the overall excess is 
positive.  By contrast, when the source star is located at the 
adjacent position of the magenta circle, it encompasses the 
same number of the two large positive-excess regions but now 
four large negative-excess regions including two additional 
regions around fold caustics and one inside the caustic, 
resulting in an overall negative excess.  The other perturbation 
regions can be explained in a similar way.

The existence of the characteristic features in the pattern 
of central perturbation region provides an important diagnostic 
tool that can be used to identify the existence of a companion 
for a high-magnification event affected by severe finite-source 
effect.  This is because the light curve of an event where the 
source crosses these features will exhibit a distinctive signal 
that is characterized by short-duration perturbations of either 
positive or negative excess and a flat residual region between 
these short-duration perturbations.  We present an example light 
curve of such an event in Figure~\ref{fig:three}.  Recently, a 
planetary-lensing event exhibiting a similar planetary signal 
was actually detected (MOA-2007-BLG-400: Dong et al.\ 2008).  
Considering that high-magnification events are prime targets 
of the current microlensing follow-up observations, we predict 
that more of such planetary signals would be detected.

We note, however, that the existence of these characteristic 
signals does not necessarily confirm the existence of a 
planetary companion.  This is because such signals can also 
be produced by a wide ($s\gg 1.0$) or a close ($s\ll 1.0$) 
binary companions that can also give rise to small central
caustics.  Therefore, detailed modeling should be done for 
the complete characterization of the lens system.

\section{Summary and Conclusion}

We investigated the effect of a finite source on the central 
perturbation pattern. From this, we found that the dependency 
of the finite-source effect on the caustic shape is weak and 
perturbations survive even when the source is substantially 
bigger than the caustic.  Specifically, we found that 
perturbations with fractional magnification excess $\geq 5\%$ 
survive when the source star is roughly 4 times bigger than 
the caustic.  We also found characteristic features that 
commonly appear in the perturbation patterns of lens systems 
affected by severe finite-source effect and thus can be used 
for the diagnosis of the existence of a companion.  These 
features form in and around a circle with its center 
located at the caustic center and a radius corresponding to 
that of the source star.  The light curve of an event where 
the source crosses these features will exhibit a distinctive 
signal that is characterized by short-duration perturbations 
of either positive or negative excess and a flat residual 
region between these short-duration perturbations.

\acknowledgments 
This work was supported by the Astrophysical Research Center 
for the Structure and Evolution of the Cosmos (ARCSEC) of 
Korea Science and Engineering Foundation (KOSEF) through 
Science Research Center (SRC) program.

\end{document}